\documentstyle[12pt,epsf]{article}
\begin{document}
\begin{flushright} 
ACT-2/98 \\
CTP-TAMU-8/98 \\
OUTP-98-19P \\
quant-ph/9803005 \\
\end{flushright}
\begin{center}
\Large{\bf Ferroelectrics and their possible involvement in biology }
\vskip1cm
{\sf \large N.E. Mavromatos $^{1~}$,  D.V. Nanopoulos $^{2~}$,
 I. Samaras $^{3}$,~ K. Zioutas $^{3~}$.}
\end{center}

\vskip1.0cm

\noindent
1) P.P.A.R.C. Advanced Fellow, University of Oxford, Department of 
Physics, Theoretical Physics,
1 Keble Road, OX1 3NP, U.K. 

\noindent
2) Department of Physics, Texas A \& M University, 
College Station, TX 77843-4242, USA, and 
Astroparticle Physics Group, Houston
Advanced Research Center (HARC), The Mitchell Campus,
Woodlands, TX 77381, USA, and Academy of Athens, Chair of Theoretical Physics, 
Division of Natural Sciences, 28 Panepistimiou Avenue, 
Athens 10679, Greece.

\noindent
3) Physics Department, University of Thessaloniki, GR54006 Thessaloniki,
Greece.

\begin{center}
\vskip1.0cm
{\it 27. 2. 1998 }
\vskip1.0cm
\section*{Abstract}
\end{center}
{\sf We present the main properties of ferroelectricity, with emphasis
given to a specific family of hydrated ferroelectric crystals, which can serve
as model systems for corresponding configurations in biology like the microtubules.
An experimental method is described, which allows to establish the ferroelectric
property of microtubules in suspension.

}
\vskip2.5cm

\begin{center}
{\large \sf \it Presented at the Workshop on the Molecular Biophysics
of the Cytoskeleton, August 18-22, 1997, Banff, Alberta, Canada.}
\end{center}

\vskip1.5cm

\newpage

\section*{1. Introduction}

Thales of Miletus (6th century B.C.) made the first observations of
electricity by rubbing pieces of amber being biological in origin,
thanks to macroscopic charge
separation. Spontaneous charge separation appears also with ferroelectric
materials, where at the molecular level the positive charge center is shifted
relative to the negative charge distribution. The perfect alignment
of these individual electric dipoles gives rise to the ferroelectricity.
Thus, the ferroelectric state exhibits spontaneously a macroscopic
electric dipole moment even in the absence of an external electric field.
This occurs below a characteristic temperature, the Curie temperature T$_C$,
where the individual electric dipoles  of the molecules get spontaneously
aligned; for comparison, to polarize a dielectric material to a similar value,
one needs electric fields of the order MV/cm to GV/cm, and this distinguishes
ferroelectricity.

The consequence of the coherent single dipole alignment is the appearance of
unpaired positive and negative bound charges on the two end-faces along the
direction of the spontaneous ferroelectric polarization ${\bf P_s}$.
In the ideal case, a piece of ferroelectric material is equivalent to an
empty capacitor of the same dimensions, each plate being charged with
${\sf Q_{\pm}}=\pm {\bf P_s}$.
 Assuming
$\mid {\bf P_s}\mid$ values by factor $\sim1000$ below those encountered often with
crystals, the associated surface charge density ($\sigma$) can still be as much
as $\sigma \approx 10^{11}$ charges/cm$^2\equiv 10^3$ charges/$\mu$m$^2$.
The maximum electric field strength inside matter is given by
${\bf E=-P_s/}\epsilon$, with $\epsilon$ being the dielectric constant of
the medium. For example, for $\mid{\bf P_s}\mid$=1 mC/m$^2$ only, the associated
electric  field  is of the order of $\epsilon ^{-1}\cdot $MV/cm.
Such electric fields exist only transiently, and,
they can give rise even to the self-emission of energetic electrons
($\sim 100$ keV electrons have been measured (s. Gundel et al., 1989));
depending on the actual conditions those electrons might ionize the
surrounding, or, they can  cause other related effects.

In this section, we present in short the main properties of ferroelectricity,
a phenomenon which is also described in the text books, but we also refer to
a few related articles, which can be read actually by non-specialists
(Zioutas $\&$ Daskaloyannis, 1985, Gundel et al., 1989, Riege, 1994), or,
modern theoretical ones (Resta, 1997).

It is worth remembering that ferroelectricity and ferromagnetism have at first
sight many common properties. Comparing both types of materials, the main
difference  comes from the (at least practically) non-existence of magnetic
monopoles. The ubiquitous electric charges compensate the large polarization
electric fields, and, they determine the transient ferroelectric behaviour
during polarization change.  
Thus, the ferroelectric state is completely described by its widely known static
properties as well as its dynamical behaviour associated with a temperature
change, or, applying externally pressure and/or electric pulses. In fact, every
ferroelectric material is also piezoelectric and pyroelectric, but not
{\it vice versa}.

Obviously, any change of the intrinsic spontaneous ferroelectric
polarization (${\bf \Delta P_s})$ is associated with a corresponding surface
charge change (${\sf \Delta Q_s}$). These
bound polarization charges attract (either from the surrounding or from the
ferroelectric medium itself) an equal number of charges,
but opposite in sign, since no environment can stand those electric fields.
In short, a ferroelectric in its static state consists of a polarized crystal
or polycrystalline medium, and, quasi free screening surface charges,
which partly migrate into the bulk of the ferroelectric.
The time period needed, for this charge neutralization to be completed,
defines the dynamical behaviour of ferroelectrics.
In general, this transient time is much longer than the preceding switching
time. Depending on the applying
electric field, temperature and the ferroelctric compound itself, transient
times can be in the $\sim \mu$sec to sec region.
Many ferroelectric properties during polarization change appear
much more pronounced compared with those we know in the static case. For
example, the dielectric constant takes enormous values around the phase
transition temperature.

In fact, a ferroelectric sample in its static state might remain unnoticed at
first sight. It is only the change of its spontaneous polarization vector
(${\bf P_s}$), which makes actually ferroelectricity visible. We mention here a few mostly
encountered  cases when working with such materials.
~~{\bf a)}~ Let the temperature increase from the ferro- to the paraelectric
state. During the phase transition, the intrinsic polarization 
disappears, and, one is left suddenly with the unpaired screening charges on
each end-surface, which must be screened so to say, because of the associated
strong electric fields. Since those charges are quasi-free, the result will
be : either a current flow through the bulk of the ferroelectric material with
both charge polarities compensating each other, or, for sufficient high
intrinsic resistance of the sample, those almost unbound charges will be
removed from each end-surface by their own repulsive Coulomb force
(i.e. self-emission).
~~{\bf b)}~ In the opposite process, i.e. by lowering the temperature from the
paraelectric to the ferroelectric state, the onset of the spontaneous
polarization attracts screening charges from the surrounding, because usually
those transient electric fields are far above the electrical breakdown limit
of any material, including the ferroelectric sample itself, if one should
consider this to happen in space.
~~{\bf c)}~ Reversal of the spontaneous polarization takes place, for example,
by applying an external electric field ($\sim$kV/cm is usually sufficient),
being either alternating or pulsed. Thus, during a hysteresis loop measurement
two polarization reversals occur per cycle.
In a complete reversal of the intrinsic spontaneous polarization, the new
bound polarization charges on each end-face of the sample have the
same polarity as the previous screening charges. This makes the associated
transient electric fields even stronger than those appearing in the first two
cases mentioned before. The result is finally  a complete rearrangement of the
old  screening charges, giving rise to a more pronounced dynamical behaviour,
depending on the conductivity of the sample and other factors.

Finally, we mention in short a few other known effects, which occur when
changing the polarization vector by ${\bf \Delta P_s}$, or, much more enhanced
during phase transition, which afterall can be of relevance for similar
phenomena with the predicted {\it biological} ferroelectrics. These are : the flow of
pyroelectric current through the sample, which occasionally takes the form of
spikes (Barkhausen pulses), similarly to the self-emission of energetic
electrons, or, bursts of microwave radiation 
(Caspers, Riege $\&$ Zioutas, 1988).
We also like to mention here the antiferroelectric state, which is
possibly of not minor importance. This is a peculiar polarized state, since it
consists of quasi infinite many 180$^o$ domains of opposite alignment each,
which does not require screening charges. However, the transition 
ferroelectric$\large{\rightleftharpoons}$antiferroelectric gives also rise  to a
similar dynamical behaviour as for the mentioned
paraelectric$\large{\rightleftharpoons}$ferroelectric phase transition.

\section*{2. Hydrated ferroelectric compounds}

We like to focus now on a very specific family of hydrated ferroelectric
crystals, with the widely investigated potassium ferrocyanide trihydrate
(K$_4$Fe(CN)$_6\cdot$3H$_2$O), in short {\sf KFCT}
(Helwig, Kl\"opperpieper, $\&$ M\"user, 1978),
taken to be a representative case.
We think that this type of ferroelectricity might serve
as a model configuration in biology, with emphasis given to the microtubules,
which no doubt play a fundamental r\^ole in cell's functioning.
In fact, these subcellular constituents have attracted the recent years the
particular interest of biophysicists. 

Microtubules have
been predicted to posses oriented dipoles, assuming ferroelectric phase, which
appears to be optimal for microtubule signaling and assembly/disassembly
(Sataric et al., 1993,  Tuszynski et al., 1995).
Such  processes are obviously of basic importance. 
In the light of previous work (Mavromatos and Nanopoulos 1997,1998),
relating the ferroelectric properties to the appearance of 
mesoscopic quantum-coherent states in internal regions of 
the microtubule cylinders, playing the r\^ole of (thermally isolated) 
electromagnetic cavities, ferroelectricity may also be important 
for quantum mechanical aspects of microtubular arrangements
in the cell. 
Specifically, it was suggested 
in [Mavromatos and Nanopoulos 1998] that 
ferroelectricity in the dimer medium, strengthens isolation of the   
`cavity regions', thereby leading to a larger probability for the coherent 
states to form and live long enough, so as to transport information 
and energy across the microtubular arrangement. The r\^ole of the 
surrounding 
water molecules for the ferroelectricity in the dimer medium was emphasized.

In this latter respect we now mention that 
the interesting property of {\sf KFCT} is the origin of its ferroelectricty.
In fact, the alignment of the water molecules by the surrounding crystalline
structure ($!$) causes the measured macroscopic spontaneous polarization
$\mid {\bf P_s}\mid = 24.6~$mC/m$^2$, which is compatible with the known
microscopic water dipole configuration in  {\sf KFCT}
(Helwig, Kl\"opperpieper, $\&$ M\"user, 1978).
To be more specific, the onset of ferroelectric behaviour is connected with the
ordering of the hydrogen-bonded water-molecules
(Taylor, M\"uller, Hitterman, 1970).
Thus, one is tempting to conclude from this particular ferroelectric crystal
that a similar mechanism might be at work with the microtubules. 

Indeed,  
a typical microtubular arrangement  
consists of a quasi solid cylinder of 
tubulin dimers, which are arranged in 13 arrays (protofilaments). 
The interior of the microtubules, of cross-section 
diameter $14~nm$, is 
filled with water molecules, being ubiquitous in biology.
The dimers (of extent $8~nm$ in the direction of 
the protofilament axis) 
are believed to have unpaired (mobile) electric charges,
which lead to electric dipole moments (Sataric {\it et al.}, 1993, 
Tuszynski {\it et al.}, 1995). In (Mavromatos and Nanopoulos, 1997, 1998) 
it was further conjectured that there are thin layers
in the interior of the cylinders, of a thickness ranging up to few atomic 
scales (Angstr\"oms), fairly isolated from their environment,
which operate as cavities sustaining
mesoscopic quantum coherent modes. In this scenario, 
the importance of the water
molecules lies in the provision of the necessary coupling 
with the quantized elecgromagnetic radiation in the cavity, via 
their electric dipole moments, so that: (i) coherent modes 
appear in the 
interior of the cavity (dipole quanta) (Del Giudice {\it et al.} 1988),
and (ii) a coupling exists between 
these coherent modes and the dimer dipole quantum oscillation modes
(Mavromatos, Nanopoulos, 1997). The latter leads to coherent 
solitonic modes along the protofilament dimer chains, 
which have been argued to be responsible for dissipationless energy 
transfer across the microtubule. 

In view of the existence of the afore-mentioned hydrated ferroelectric 
compounds, we may now conjecture that 
such configurations favour an alignment of the water molecules (actually, 
not
necessarily only those inside the tube). 
Then,  the sofar speculated polar
properties of microtubules, and their connection to quantum physics,  
might exist in reality.

\section*{3. Suggested experimental method}

Because of the importance of such a ferroelectric property for these
subcellular constituents,
the next step is to establish or exclude it experimentally.
A suggested sensitive experimental method to confirm the predicted
ferroelectric behaviour of microtubules is via
the four probe AC impedance spectroscopy (Z($\nu$)).
This technique is
based on the measuremnt of the AC voltage to current ratio
using 4-electrodes (Figure 1a) in a quite broad
frequency interval (the so called {\it frequency domain technique}
(Figure 1b)). The experimental set up for the AC impedance measurement 
of the microtubules in liquid suspension is shown  
in figure 2. 
This method has been recognised already as a valuable tool in the study of
similar electrochemical systems (MacDonald, 1987).
There are practical advantages of this technique compared to the more simple
and old one working in the so called {\it time domain}, which
applies either pulse or
step source signals (MacDonald, 1987, Bottelberghs, 1978).  In fact, the
expected ferroelectric contribution of the suspended microtubules to the
measured impedance Z($\nu$) can only be 
observed in the higher frequency domain (s. Figure 1b).
Otherwise, low frequency electrochemical phenomena, being certainly
present in such a liquid suspension of microtubules, complicate the whole
measuring procedure, which will suppress the appearance of a ferroelectric
component inside the sample.

In practice, working with a suspension of microtubules additional experimental
modifications might well become necessary. Thus, in addition to a conventional
DC polarization for the Z($\nu$) measurement one can also apply a pseudo-DC polarization,
in order to avoid quasi-static complex electrochemical phenomena due to
the electrode's
polarization coming from slow and heavy ion transport inside the sample.
The onset of ferroelectricity can be identified due to a resonance-like
response in the phase and the amplitude diagram of the impedance (s. Figure 1b).

\section*{Acknowledgements}

We thank J. Albers for discussions and for providing us 
with information on hydrated ferroelectrics. 
This work is based on presentations 
by N.E.M. and K. Z. at the Workshop `Biophysics of the Cytoskeleton',
Banff, Alberta, Canada, August 18-22 1997. We thank J. Tuszynski 
for his interest in our work. 
The work of N.E.M. is supported by P.P.A.R.C. (U.K.), and that 
of D.V.N. is supported
in part by D.O.E. Grant
DE-F-G03-95-ER-40917.

\newpage

\begin{center}
\section*{REFERENCES}
\end{center}

\noindent
Bottelberghs, P. H. \underline{\it Solid Electrolytes,} Edits. P. Hagenm\"uller and
W. van Gool,

Academic {\bf 1978} p 145.

\noindent
Caspers, F., Riege, H., Zioutas, K., unpublished {\bf 1988}.

\noindent Del Giudice, E.; G. Preparata, G.; Vitiello, G.
\underline{  Phys. Rev. Lett.} {\bf 61} (1988), 1085.

\noindent
Gundel, H., Riege, H., Wilson, E. J. N., Handerek, J., Zioutas, K.,
\underline{\it Nucl. Instr. and Methods}

{\bf 1989}, \underline{\it A280}, 1.

\noindent
Helwig, J., Kl\"opperpieper, A. and M\"user, H. E.,
\underline{\it Ferroelectrics} {\bf 1978} \underline{\it 18}, 257.

\noindent
MacDonald, J. R. 
\underline{\it Impedance Spectroscopy, Emphasizing Solid Materials}, (Ed.),

Wiley, New York {\bf 1987}.

\noindent
Mavromatos N.E.;  Nanopoulos, D.V., quant-ph/9708003,
\underline{  Int. J. Mod. Phys. B12}, {\bf 1997}, 

in press.

\noindent
Mavromatos, N.E., Nanopoulos, D.V., quant-ph/9802063, {\bf 1998}, 
contribution to the Proc. 
of
the Workshop \underline{Biophysics
of the Cytoskeleton} (Banff, August 18-22 1997, Canada).

\noindent
Resta, R., \underline{\it Europhysics News} {\bf 1997}, \underline{\it 28(1)},
18, and references therein.

\noindent
Riege, H., \underline{\it Nucl. Instr. and Methods}
{\bf 1994}, \underline{\it A340}, 80.

\noindent
Satari\'c, M. V., Tuszynski, J., Zakula, R. B.,  \underline{\it Phys. Rev.}
{\bf 1993} \underline{\it E48}, 589.

\noindent
Taylor, J. C., M\"uller, Hitterman,
\underline{\it Acta Cryst.} {\bf 1970} \underline{A26}, 559.

\noindent
Tuszynski, J., Hameroff, S., Satari\'c, M. V., Trpisov\'a, B., Nip, M. L. A.,

\underline{\it J. theor.  Biol.} {\bf 1995}, \underline{\it 174}, 371.

\noindent
Zioutas, K., Daskaloyannis, C.
\underline{\it Nucl. Instr. and Methods} {\bf 1985},
\underline{\it B12}, 200.

\vskip2.0cm

\newpage 
\begin{center}
\Large{\bf Figure captions}
\end{center}

~

\vskip2.0cm

\noindent
{\bf Figure 1}.~ AC impedance measurement of a suspension of microtubules using :
{\bf (a)} a 4 probe electrode configuration in order to avoid electrode
polarization phenomena, and
{\bf (b)} two main experimental regimes in the measured amplitude and phase of
the complex impedance Z($\nu$).

\vskip0.5cm

\noindent
{\bf Figure 2}.~ Experimental set-up for the AC impedance measurement of the
microtubules in liquid suspension,
using evaporated 4 Pt or Au electrical contacts on a glass substrate mounted on a
Cu temperature head, allowing measurements at different temperatures (left).
The diagram of the experimental set-up with the possibility of a pseudo
DC offset is also shown (right).

\newpage

\begin{centering}
\begin{figure}[htb]
%\epsfxsize=4in
%
%\vspace{2cm}
%

%\centerline{\rotate{\rotate{\rotate{\epsffile{fig33.ps}}}}}
\centerline{\epsffile{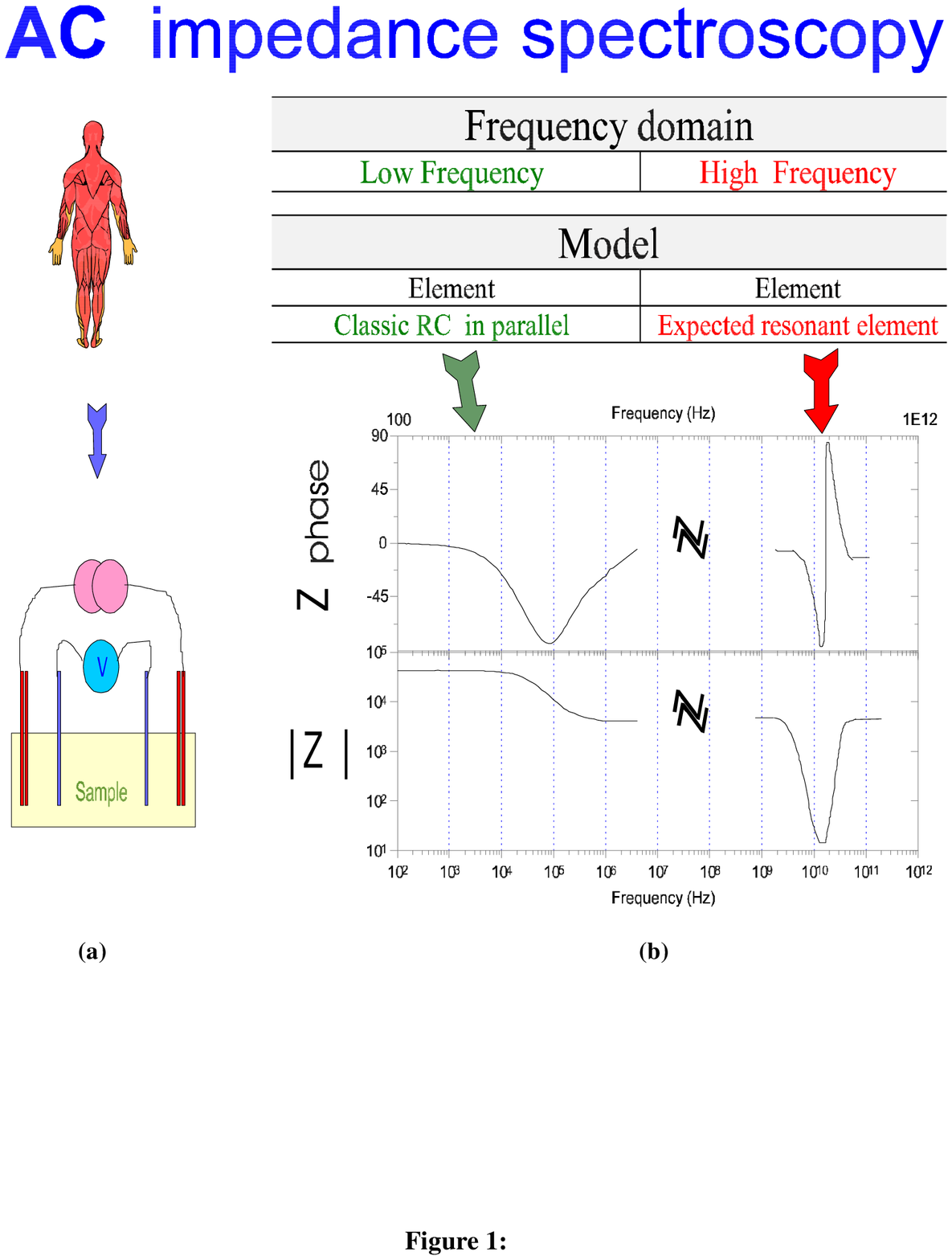}}
\vspace{1cm}
\end{figure}
\end{centering}

\end{document}